\documentclass[final,5p,times,fleqn,twocolumn]{elsarticle}

\usepackage{dcolumn}
\usepackage{subcaption}
\usepackage{hyperref}
\usepackage{amsmath}
\usepackage{amssymb}
\usepackage{amsthm}
\usepackage{mathrsfs}
\usepackage{mathtools}
\usepackage{bm}
\usepackage{float}
\usepackage{orcidlink}
\hypersetup{colorlinks=true}

\begin{document}


\begin{frontmatter}

\title{\texttt{soliton\_solver}: A GPU-based finite-difference PDE solver for topological solitons in two-dimensional non-linear field theories}

\author{Paul Leask\orcidlink{0000-0002-6012-0034}}
\ead{palea@kth.se}
\affiliation{organization={Department of Physics},
            addressline={KTH Royal Institute of Technology}, 
            city={Stockholm},
            postcode={SE-10691},
            country=Sweden}

\begin{abstract}
This paper introduces \texttt{soliton\_solver}, an open-source GPU-accelerated software package for the simulation and real-time visualization of topological solitons in two-dimensional non-linear field theories.
The software is structured around a theory-agnostic numerical core implemented using Numba CUDA kernels, while individual physical models are introduced through modular theory components.
This separation enables a single computational framework to be applied across a broad class of systems, from nanoscale magnetic spin textures in condensed matter physics to cosmic strings spanning galaxies in high energy physics.
The numerical backend provides finite-difference discretization, energy minimization, and GPU-resident evaluation of observables.
A CUDA--PyOpenGL rendering pipeline allows direct visualization of evolving field configurations without staging full arrays through host memory.
The package is distributed in Python via PyPI and supports both reproducible batch simulations and interactive exploration of metastable configurations, soliton interactions, and model-dependent initial states.
We describe the software architecture, numerical workflow, and extensibility model, and we present representative example applications.
We also outline how additional theories can be incorporated with minimal modification of the shared numerical infrastructure.
\end{abstract}


\end{frontmatter}

\section*{Code metadata}

\begin{table*}[ht]
    \centering
    \caption{Code metadata}
    \begin{tabular}{p{0.4\linewidth} p{0.55\linewidth}}
        \hline
        Current code version & 1.1.0 \\
        Permanent link to code repository used for this code version & \href{https://github.com/Paulnleask/soliton_solver}{https://github.com/Paulnleask/soliton\_solver} \\
        Permanent link to reproducible capsule & \\
        Legal Code License & \href{https://github.com/Paulnleask/soliton_solver/blob/main/LICENSE}{https://github.com/Paulnleask/soliton\_solver/blob/main/LICENSE} \\
        Code versioning system used & Git \\
        Software code languages, tools, and services used & Python, Numba, CUDA, PyOpenGL, ModernGL, glfw, cuda-python \\
        Compilation requirements, operating environments \& dependencies & Python 3.10+, NVIDIA GPU, CUDA-capable drivers, OpenGL support, NumPy, Numba, ModernGL, glfw, PyOpenGL, cuda-python \\
        If available, link to developer documentation/manual & \href{https://github.com/Paulnleask/soliton_solver/blob/main/README.md}{https://github.com/Paulnleask/soliton\_solver/blob/main/README.md} \\
        Support email for questions & palea@kth.se \\
        \hline
    \end{tabular}
\end{table*}


\section{Introduction}
\label{sec:introduction}

Topological solitons are finite-energy, spatially localized solutions of non-linear field theories that behave as quasi-particles, often representing collective emergent phenomena.
Their stability is not merely dynamical but also topological in nature: field configurations fall into distinct homotopy classes, and cannot be continuously deformed into one another.
As a result, a soliton carries a conserved topological charge, often describing the total number of constituent particles, and cannot unwind into the vacuum.

While their existence is usually guaranteed by these topological arguments, obtaining explicit solutions is rarely straightforward.
The governing Euler--Lagrange field equations are typically non-linear partial differential equations, and analytic solutions exist only in a small number of highly symmetric or integrable cases.
Even then, extracting physically relevant properties can be non-trivial.

In practice, studying topological solitons requires solving these field equations numerically, often under non-trivial boundary conditions and across different topological sectors.
This motivates the development of dedicated computational tools, designed to efficiently construct, analyse, and explore soliton solutions in a systematic way.

Topological solitons arise in a broad class of non-linear field theories, ranging from topological spin textures in nanoscale condensed matter systems up to superconducting cosmic strings spanning galaxies \cite{Manton_Sutcliffe_2004}.
While the underlying physics differs between these systems, they are unified by a common numerical approach: energy minimisation.
Existing numerical tools are typically developed with a specific model class in mind.
Micromagnetic solvers such as OOMMF \cite{Donahue_1999} and MuMax \cite{Vansteenkiste_2011,Vansteenkiste_2014} are optimized for spin systems, while Ginzburg--Landau simulations are often carried out using specialized solvers such as Svirl \cite{Svirl} and pyTDGL \cite{pyTDGL}, or within more general-purpose PDE frameworks tailored to particular order parameters or coupling structures.
This specialization is effective within a given domain, but it limits reuse across related theories and makes it more difficult to prototype new coupled systems that combine features from multiple model classes.

\texttt{soliton\_solver} was developed to address this limitation.
The package provides a reusable GPU-accelerated numerical core for two-dimensional field theories together with a modular interface through which individual models define their field content, parameters, energy densities, initialization procedures, observables, and visualization routines.
It is a Python package built around Numba CUDA kernels, a theory registry, and CUDA--PyOpenGL interoperability for interactive visualization, with distribution provided through PyPI.

The aim of the present work is therefore to document the software.
The primary contribution is a reusable computational framework consisting of a theory-agnostic finite-difference engine, a general energy minimization algorithm, a runtime theory-loading mechanism, and an interactive visualization backend.
Built-in models are used to illustrate how the framework is applied in practice and to demonstrate how new theories can be incorporated with minimal modification of the shared numerical infrastructure.


\section{Software description}
\label{sec:software_description}


\subsection{Software architecture}

The \texttt{soliton\_solver} package is structured around a clear separation between a reusable numerical backend and model-specific theory modules.
The current repository organization reflects this design, with the package divided into a core numerical engine, a collection of modular theory implementations, a visualization backend, and a set of runnable examples.
This separation allows the same GPU-based solver infrastructure to be applied across multiple non-linear field theories without modification of the numerical core.

At a high level, the package consists of:
\begin{itemize}
    \item a \texttt{core/} layer implementing finite-difference operators, time-stepping routines, simulation drivers, and GPU memory management,
    \item a \texttt{theories/} layer in which each model specifies its field content, energy functional, parameters, initialization procedures, and optional visualization routines,
    \item a \texttt{visualization/} layer providing CUDA--PyOpenGL rendering for interactive inspection of evolving field configurations,
    \item an \texttt{examples/} layer containing runnable demonstrations.
\end{itemize}

A central architectural component is the theory registry together with a dependency-injection pattern.
Rather than embedding a fixed physical model into the solver, \texttt{soliton\_solver} loads a selected theory at runtime and combines it with a generic \texttt{Simulation} driver.
In this way, discretization, minimization, reductions, and rendering remain independent of the specific model, while theory-dependent logic is localized within dedicated modules.


\subsection{Software functionalities}

The package provides a set of core capabilities that are shared across all supported theories:
\begin{itemize}
    \item GPU-accelerated finite-difference simulation of non-linear partial differential equations in two spatial dimensions,
    \item an accelerated gradient-based minimization for relaxation to static configurations,
    \item real-time CUDA--PyOpenGL visualization of evolving field configurations,
    \item a modular interface for the implementation of new field theories,
    \item example models spanning gauge theories, magnetic systems, condensates, and coupled superconducting--magnetic systems.
\end{itemize}

These features are documented in the repository and are implemented in a way that is independent of the specific physical model.
The current implementation targets NVIDIA GPUs through Numba CUDA, with the runtime stack consisting of Python, CUDA, OpenGL, and associated libraries such as glfw, PyOpenGL, and \texttt{cuda-python}.


\subsection{Supported models and modular theory interface}

A key design choice in \texttt{soliton\_solver} is that physical models are implemented as theory modules that are registered and loaded at runtime rather than hard-coded into the solver.
The registry stores metadata such as the canonical theory name, import path, version, and aliases, while the shared \texttt{Simulation} class binds the selected theory to the common numerical workflow.

At runtime, a theory module supplies model-specific components, including field definitions, parameter packing, initialization routines, CUDA kernels, and optional visualization helpers.
The core package provides grid construction, memory allocation, derivative evaluation, time integration, and reduction operations.
This separation ensures that the same simulation driver can be used across different classes of models without modification.

The software currently includes built-in theories describing:
\begin{itemize}
    \item abelian Higgs cosmic strings \cite{Nielsen_1973,Ruback_1988,Hindmarsh_1995,Hindmarsh_2009} and vortices in Ginzburg--Landau superconductors \cite{Berger_1989,Ginzburg_2009};
    \item vortex anyons in topologically ordered superconductors \cite{Hansson_2004,Leask_2026};
    \item superfluid vortices in rotating, trapped Bose--Einstein condensates \cite{Fetter_2009,Nitta_2014,Chen_2023};
    \item skyrmions in chiral liquid crystals \cite{Bogdanov_2003,Leonov_2014,Duzgun_2018,Ackerman_2014,Selinger_2017}, including the effects of flexoelectric depolarization \cite{Leask_2025};
    \item composite magnetic skyrmion-superconducting vortex states \cite{Nothhelfer_2022,Mukherjee_2025,Leask_Ross_Babaev_2026} in ferromagnetic superconductors \cite{Varma_1979,Varma_1981};
    \item fractional vortices and skyrmions in anisotropic Ginzburg--Landau superconductors  \cite{Babev_Speight_Winyard_2023,Babev_Winyard_2023,Speight_Winyard,Zhang_2020,Talkachov_Leask_Babaev_2026};
    \item skyrmions in chiral magnets \cite{Bogdanov_1994,Fert_2017,Rybakov_2019,Kuchkin_2020}, including the effects of demagnetization \cite{Leask_Speight_2026,Fratta_2020};
    \item and baby skyrmions in the two-dimensional Skyrme model with various potentials \cite{Weidig_1999,Leask_2022}, such as the standard \cite{Piette_1995}, easy plane \cite{Jaykka_2010,Nitta_2013}, broken \cite{Jaykka_2012,Winyard_2013}, dihedral \cite{Ward_2004}, and aloof \cite{Salmi_2014}.
\end{itemize}
These models span vastly different length and energy scales, but they all reuse the same GPU stencil machinery and simulation driver.
A summary of the currently supported theories (including the physical system(s) they describe, their representative fields, and associated energy functionals) is shown in Tab. \ref{tab: Theories}.

This architecture enables reuse beyond a single application domain.
Rather than providing a solver for a specific field theory, \texttt{soliton\_solver} offers a general computational framework in which new models can be introduced by implementing a compact theory module that satisfies the registry interface.

\begin{table*}[t]
    \centering
    \caption{Currently supported two-dimensional theories alongside their corresponding soliton type(s), field(s) and associated energy functional.}
    \begin{tabular}{|m{2.7cm}|m{1.4cm}|m{1.15cm}|m{10.5cm}|}
        \hline
        Theory \newline \emph{Physical system} & Solitons & Fields & Energy Functional \\
        \hline\hline
        Abelian Higgs / \newline Ginzburg-Landau \newline \emph{Cosmic strings /} \newline \emph{superconductors} & Vortices & $\psi \in \mathbb{C}$, $\vec{A} \in \mathbb{R}^2$ & \begin{align*}
            E_{\textup{AH}}[\psi, \vec{A}] =  \int_{\mathbb{R}^2} \textup{d}^2x  \left\{\frac{1}{2}|\vec{D}\psi|^2 + \frac{1}{2}|\vec{\nabla}\times\vec{A}|^2 + \frac{\lambda}{8} \left( u^2 - |\psi|^2 \right)^2 \right\}
        \end{align*} \\
        \hline
        Baby Skyrme \newline \emph{Planar analogue of nuclear skyrmions} & Skyrmions & $\vec{m} \in \mathbb{R}^3$ & \begin{align*}
            E_{\textup{BS}}[\vec{m}] = \int_{\mathbb{R}^2} \textup{d}^2x  \left\{\frac{1}{2} |\nabla \vec{m}|^2 + \frac{\kappa^2}{4} \left| \partial_i \vec{m} \times \partial_j \vec{m} \right|^2+ V(\vec{m}) \right\}
        \end{align*} \\
        \hline
        Bose-Einstein \newline condensate \newline \emph{Rotating super-} \newline \emph{fluids} & Vortices & $\Psi \in \mathbb{C}$ & \begin{align*}
            E_{\textup{BEC}}[\Psi] = \int_{\mathbb{R}^2} \textup{d}^2x  \left\{ \frac{\hbar^2}{2m}|\vec{\nabla} \Psi|^2 + \frac{1}{2}m \omega^2 |\vec{r}|^2|\Psi|^2 + \frac{g}{2}|\Psi|^4 - \Omega \Psi^* \hat{L}_z \Psi \right\} \\
            \hat{L}_z = -i\hbar \left( x \partial_y - y \partial_x \right)
        \end{align*} \\
        \hline
        Chern-Simons-Landau-Ginzburg \newline \emph{Anyon super-}\newline \emph{conductors} & Anyons & $\psi \in \mathbb{C}$, $\vec{A} \in \mathbb{R}^2$, $A_0\in\mathbb{R}$ & \begin{align*}
            E_{\textup{CSLG}}[\psi,\vec{A}] = E_{\textup{AH}}[\psi,\vec{A}] + \int_{\mathbb{R}^2} \textup{d}^2x  \left\{ \frac{1}{2}|\vec{\nabla} A_0|^2 + \frac{1}{2}q^2|\psi|^2 A_0^2 \right\} \\
            \left(-\nabla^2 +q^2|\psi|^2\right)A_0 = -\kappa (\partial_1 A_2 - \partial_2 A_1)
        \end{align*} \\
        \hline
        Magnetization \newline extended Ginzburg-Landau \newline \emph{Ferromagnetic} \newline \emph{superconductors} & Vortices, Skyrmions & $\vec{m} \in \mathbb{R}^3$, $\psi \in \mathbb{C}$, $\vec{A} \in \mathbb{R}^3$ & \begin{align*}
            E_{\textup{FS}}[\vec{m}, \psi, \vec{A}] = \int_{\mathbb{R}^2} \textup{d}^2x \left\{ \frac{1}{2}|\vec{D}\psi|^2 + \frac{1}{2}|\vec{\nabla}\times\vec{A}|^2 + \frac{\gamma^2}{2}|\nabla \vec{m}|^2 - \vec{m} \cdot (\vec{\nabla} \times \vec{A})  \right. \\ \left. + \eta_1|\vec{m}|^2|\psi|^2 + \eta_2 |\nabla\vec{m}|^2|\psi|^2+\frac{a}{2}|\psi|^2 + \frac{b}{4}|\psi|^4 + \frac{\alpha}{2}|\vec{m}|^2 + \frac{\beta}{4}|\vec{m}|^4 \right\}
        \end{align*} \\
        \hline 
        Micromagnetic \newline \emph{Chiral magnets} & Bimerons, Skyrmions & $\vec{n} \in \mathbb{R}^3$, $\phi \in \mathbb{R}$ & \begin{align*}
            E_{\textup{CM}}[\vec{n}] = \int_{\mathbb{R}^2}\textup{d}^2x \left\{ \frac{J}{2}|\textup{d}\vec{n}|^2 + \mathcal{D} \,\vec{d}_i\cdot(\vec{n}\times\partial_i\vec{n}) + M_sV(\vec{n}) + \frac{1}{2\mu_0}  |\vec{\nabla}\phi|^2 \right\} \\
            -\nabla^2 \phi = -\mu_0\vec{\nabla}\cdot(M_s\vec{n})
        \end{align*} \\
        \hline
        Multicomponent Ginzburg-Landau \newline \emph{Anisotropic $s$+$id$} \newline \emph{superconductors} & Fractional vortices, \newline Skyrmions & $\Delta_s \in \mathbb{C}$, $\Delta_d \in \mathbb{C}$, $\vec{A} \in \mathbb{R}^2$ & \begin{align*}
            E_{s+id}[\Delta_s, \Delta_d,\vec{A}] = \int_{\mathbb{R}^2} \textup{d}^2x  \left\{ \frac{1}{2} \gamma_{jk}^{\alpha\beta} (D_j \Delta_\alpha)^* D_k \Delta_\beta + \frac{1}{2}|\vec{\nabla}\times\vec{A}|^2 + \alpha_1 |\Delta_s|^2 \right. \\ \left.  + \alpha_2 |\Delta_d|^2 + \beta_1 |\Delta_s|^4 + \beta_2 |\Delta_d|^4 + \beta_3 |\Delta_s|^2 |\Delta_d|^2 + \beta_4 \left( \Delta_s^2 \bar{\Delta}_d^2 + \bar{\Delta}_s^2 \Delta_d^2 \right) \right\}
        \end{align*} \\
        \hline
        Oseen-Frank \newline \emph{Chiral liquid} \newline \emph{crystals} & Merons, Skyrmions & $\vec{n} \in \mathbb{R}^3$, $\phi \in \mathbb{R}$ & \begin{align*}
            E_{\textup{LC}}[\vec{n}] = \int_{\mathbb{R}^2} \textup{d}^2x  \left\{ \frac{K}{2} |\nabla \vec{n}|^2 + K q_0 \, \vec{d}_i\cdot(\vec{n}\times\partial_i\vec{n}) + V(\vec{n}) + \frac{\varepsilon_0}{2} |\vec{\nabla} \phi|^2 \right\} \\ -\nabla^2 \phi = -\frac{1}{\varepsilon_0}\vec{\nabla} \cdot \vec{P}_f[\vec{n}], \quad \vec{P}_f[\vec{n}] = e_1 \left[ (\vec{\nabla} \cdot \vec{n}) \vec{n} \right] + e_3 \left[ \vec{n} \times (\vec{\nabla} \times \vec{n}) \right]
        \end{align*} \\
        \hline\hline
    \end{tabular}
    \label{tab: Theories}
\end{table*}


\section{Numerical workflow}
\label{sec:numerical_workflow}


\subsection{Shared discretization and solver core}

The numerical core of \texttt{soliton\_solver} operates on a rectangular two-dimensional lattice with a fixed halo width associated with the finite-difference stencils.
A shared parameter system defines the lattice dimensions, grid spacings, number of stored field components, and solver time step, after which theory-specific parameters are packed into integer and floating-point device arrays.

The default configuration employs a two-dimensional grid with fourth-order finite-difference stencils and halo width two.
Let lattice spacing in the $x$-direction be $\Delta x$.
Then the finite difference approximation of the first derivative of a field component $\phi$ in the $x$-direction is
\begin{equation}
    \partial_{x}\phi_{i,j} \approx \frac{-\phi_{i+2,j} + 8\phi_{i+1,j} - 8\phi_{i-1,j} + \phi_{i-2,j}}{12(\Delta x)},
\end{equation}
and likewise for the second derivative,
\begin{equation}
    \partial_{x}^{2}\phi_{i,j} \approx \frac{-\phi_{i+2,j} + 16\phi_{i+1,j} - 30\phi_{i,j} + 16\phi_{i-1,j} - \phi_{i-2,j}}{12(\Delta x)^2},
\end{equation}
with analogous expressions used in the $y$ direction.
The solver time step $\Delta t$ is chosen relative to the lattice spacing $\Delta x$ through a Courant-like number, $C = \Delta t/ \Delta x$.
For stability, $C<1$ in general and is set by default to $C=0.5$.
These parameters are resolved prior to allocation of device memory, ensuring that all subsequent operations are defined in a consistent layout across theories.

The \texttt{Simulation} class then allocates a common set of device buffers, including storage for the field variables, fictitious velocity, energy gradient, derivative work arrays, Runge--Kutta stages, scalar energy buffers, and reduction buffers.
Theory-specific CUDA kernels are introduced into this shared workflow via dependency injection through the selected theory module.
As a result, the same simulation driver can be used to evolve various different physical systems without modification of the solver core.


\subsection{Energy minimization}

Topological soliton configurations are obtained by minimizing a discretized energy functional $E_h[\phi]$.
A straightforward gradient descent in $\phi$ is often robust but can be prohibitively slow because the energy landscape for solitons is typically stiff: short-wavelength modes relax quickly while long-wavelength modes (and collective coordinates such as soliton separations and internal phases) relax slowly.
An efficient alternative implemented in \texttt{soliton\_solver} is arrested Newton flow, which accelerates descent by introducing a fictitious second-order dynamics and then removing kinetic energy whenever the trajectory would climb in energy.
This method is used as an efficient minimization strategy for many soliton systems.

Concretely, we introduce a fictitious time $t$ and evolve $\phi(t)$ by
\begin{equation}
    \ddot{\phi}(t) = -\vec{\nabla}_\phi E_{h}[\phi(t)].
\label{eq: Arrested Newton flow}
\end{equation}
The initial condition is typically from rest, $\dot{\phi}(0)=0$, with $\phi(0)$ chosen to encode the desired topology.
The system \eqref{eq: Arrested Newton flow} is then integrated numerically by rewriting it as a first-order system for $(\phi,\dot{\phi})$ and applying a standard explicit time integrator such as fourth-order Runge--Kutta method with time step $\Delta t$.
The distinctive ingredient is the flow arrest: after each trial step, we compare the discrete energies, $E_h[\phi^{n+1},\Phi^{n+1}]$ and $E_h[\phi^{n}]$, and if the energy increases,
\begin{equation}
    E_h[\phi^{n+1}] > E_h[\phi^{n}],
\end{equation}
we arrest the flow by setting the velocity to zero, $\dot{\phi}^{\,n+1} \leftarrow 0$.
We then continue the evolution from the new configuration $\phi^{n+1}$ but without kinetic energy.
Intuitively, the second-order dynamics accelerates motion along shallow directions of the landscape, while the arrest criterion prevents overshooting that would otherwise produce persistent oscillations around a minimum.
This combination is typically much faster than first-order descent for multi-soliton relaxation problems.

\begin{figure*}[ht]
\centering
\includegraphics[width=0.7\textwidth]{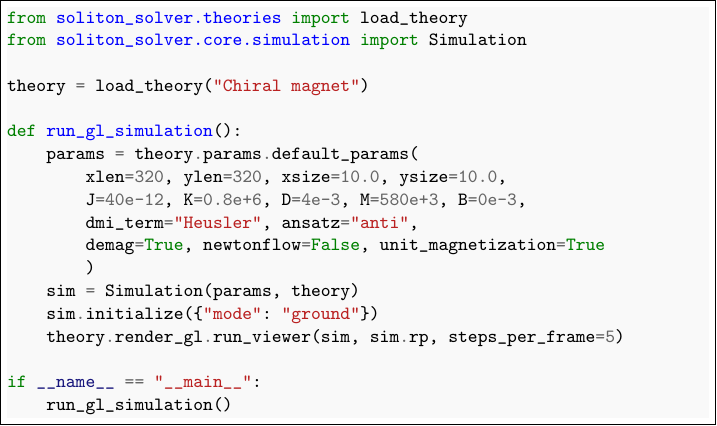}
\caption{Typical usage of \texttt{soliton\_solver} for a micromagnetic simulation of anti-skyrmions in Heusler compounds, modelled by a chiral ferromagnet with the Dzyaloshinskii--Moriya interaction (DMI), and including the effects of demagnetization.}
\label{fig: Usage}
\end{figure*}

The implementation also supports theory-dependent constraint projections.
For example, in magnetization-based models, a projection step can enforce unit-length constraints during each Runge--Kutta stage.
This separation between the shared integration scheme and model-specific constraints contributes to the generality of the solver across different classes of field theories.


\subsection{GPU reductions and interactive rendering}

In addition to pointwise field updates, the solver requires evaluation of global quantities such as total energy, convergence norms, and extrema used for visualization.
These quantities are computed using parallel GPU reductions, followed by small host-side accumulations over blockwise partial results.
This approach avoids transferring full field arrays to host memory and maintains efficient device-resident execution.

For visualization, \texttt{soliton\_solver} employs CUDA--PyOpenGL interoperability.
An OpenGL buffer is registered with CUDA and mapped into device address space for each frame.
Visualization kernels then write image data directly into this buffer, which is displayed without staging through host memory \cite{Storti_2015}.
This enables real-time inspection of evolving configurations, including energy density, order-parameter magnitude, magnetic flux density, and other observables, while the simulation remains entirely GPU resident.


\subsection{Extending the solver to new theories}

A new theory is introduced by defining its field variables, energy density, parameter set, initialization routines, and optional visualization helpers.
Because the finite-difference operators, relaxation engine, and rendering infrastructure are already provided by the core package, extending the software to a new model generally requires changes only within the corresponding theory module.
This extensibility is one of the main strengths of the package, providing an easily extendible interface.


\section{Illustrative examples and workflow}
\label{sec:examples}


\subsection{Installation and typical usage}

\texttt{soliton\_solver} can be installed either from PyPI or directly from source.
The public repository documents a standard workflow in which a theory is loaded by name, a corresponding parameter set is constructed, a \texttt{Simulation} object is initialized, and the system is then evolved through an interactive visualization environment.
The package requires Python 3.10 or newer together with an NVIDIA GPU, CUDA-compatible drivers, and OpenGL support.

A typical interactive workflow proceeds as follows:
\begin{enumerate}
    \item load a theory from the registry;
    \item construct the theory specific parameters using \texttt{default\_params(...)} with optional overrides;
    \item initialize \texttt{Simulation(params, theory)};
    \item initialize the field configuration, for example from a ground state or a model-specific ansatz;
    \item launch the interactive viewer and run minimization.
\end{enumerate}
This workflow is independent of the specific theory, reflecting the separation between the shared numerical core and the model-dependent components.


\subsection{Chiral magnet with the Dzyaloshinskii--Moriya interaction and demagnetization}

\begin{figure*}[ht]
    \centering
    \includegraphics[width=\textwidth]{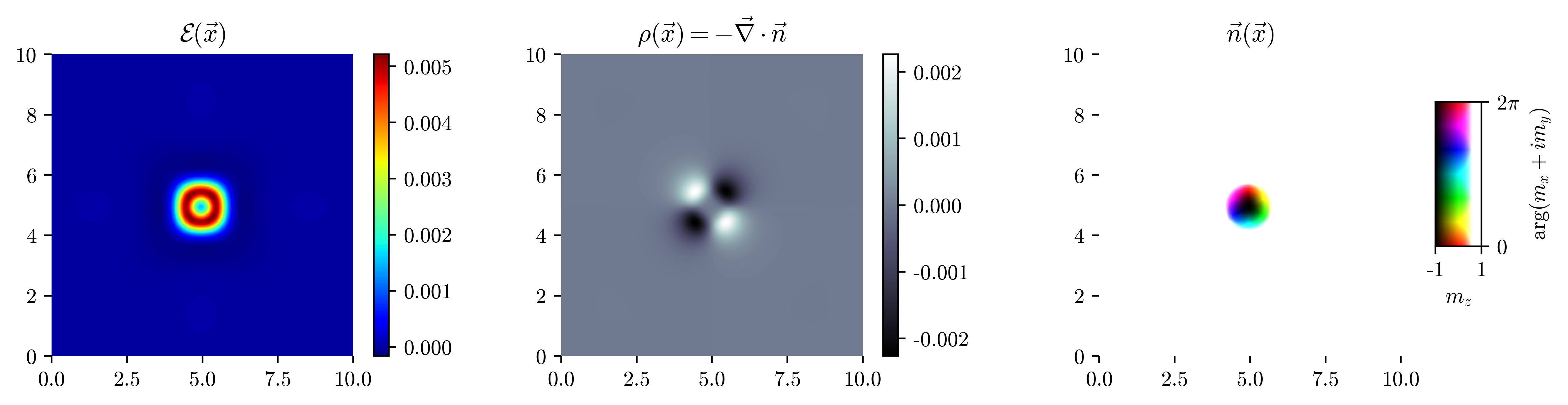}
    \caption{A single anti-skyrmion in the bulk of a Heusler compound, with the associated Dzyaloshinskii--Moriya interaction and backreaction of the demagnetization field. All panels are in dimensionless units. The left panel shows the energy density $\mathcal{E}(\vec{x})$, the middle shows the magnetic charge density $\rho(\vec{x})$, and the right shows the unit magnetization $\vec{n}(\vec{x})$. The parameters used to generate this anti-skyrmion are the ones specified in the typical usage in Fig.~\ref{fig: Usage}. These are $J=40\,\textup{pJm}^{-1}$, $\mathcal{D}=4\,\textup{mJm}^{-2}$, $M_s=580\,\textup{kAm}^{-1}$, $K_m=0.8\,\textup{MJm}^{-3}$, and $B_{\textup{ext}}=0\,\textup{T}$.}
    \label{fig: Chiral magnet}
\end{figure*}

A typical example is the chiral magnet module, which serves as a primary demonstration within the repository, with representative usage shown in Fig.~\ref{fig: Usage}. 
The code illustrates the minimal workflow: loading the theory via \texttt{load\_theory("Chiral magnet")}, specifying a physically meaningful parameter set, and launching an interactive simulation through the \texttt{Simulation} interface.

The model describes a magnetization field $\vec{n} : \mathbb{R}^2 \to S^2$ together with a scalar magnetostatic potential $\phi \in \mathbb{R}$. 
The associated energy functional is given by \cite{Leask_Speight_2026}
\begin{align}
    E_{\textup{CM}}[\vec{n}] = \int_{\mathbb{R}^2}\textup{d}^2x \left\{
    \frac{J}{2}|\nabla\vec{n}|^2 + \mathcal{D}\sum_{i=1}^3 \vec{d}_i \cdot (\vec{n} \times \partial_i \vec{n}) \right. \nonumber \\
    \left. + K_m(1 - n_3^2) + M_s B_{\textup{ext}}(1 - n_3) + \frac{1}{2\mu_0}|\vec{\nabla}\phi|^2 \right\},
\end{align}
where the magnetostatic potential is determined self-consistently through the Poisson equation
\begin{equation}
    -\nabla^2 \phi = -\mu_0 \nabla \cdot (M_s \vec{n}).
\end{equation}

The parameter set in Fig.~\ref{fig: Usage} corresponds to a typical micromagnetic regime, with exchange stiffness $J$, anisotropy constant $K_m$, DMI strength $\mathcal{D}$, saturation magnetization $M_s$, applied field $B_{\textup{ext}}$, and vacuum permeability $\mu_0$. 
The choice \texttt{dmi\_term="Heusler"} together with \texttt{ansatz="anti"} selects the antiskyrmion sector, while \texttt{demag=True} enables the magnetostatic self-interaction through the Poisson equation, so that the scalar potential $\phi$ is solved self-consistently alongside $\vec{n}$.

The variational structure therefore includes exchange, anisotropy, Zeeman, and DMI contributions, together with the nonlocal dipole--dipole interaction. 
In contrast to the purely local model, the inclusion of demagnetization breaks the degeneracy between different DMI realizations and has a direct impact on the structure and stability of the resulting textures.

This example demonstrates the ability of the solver to handle constrained vector fields ($|\vec{n}|=1$), incorporate multiple competing interactions, and couple local and nonlocal terms within a single computational framework. 
The interactive viewer shown in Fig.~\ref{fig: Usage} then allows direct manipulation of the configuration, including skyrmion placement, control of topological charge, isorotation, and switching between visualization modes.


\subsection{Rotating trapped Bose--Einstein condensate}

Another typical example is the Bose--Einstein condensate (BEC) module, which serves as a second canonical demonstration within the repository, mirroring the workflow shown in Fig.~\ref{fig: Usage}. 
The usage pattern is identical at the interface level: the theory is loaded, parameters are specified, and the simulation is initialized and visualized through the same solver infrastructure.

The model describes a single complex scalar order parameter $\psi \in \mathbb{C}$ representing the condensate wavefunction. 
The corresponding energy functional is given by \cite{Fetter_2009}
\begin{align}
    E_{\textup{BEC}}[\psi] = \, & \int_{\mathbb{R}^2} \textup{d}^2x \left\{
    \frac{\hbar^2}{2m}|\vec{\nabla} \psi|^2 + \frac{1}{2}m \omega^2 |\vec{r}|^2 |\psi|^2 + \frac{g}{2}|\psi|^4 \right\} \nonumber \\
    \, & -\Omega \int_{\mathbb{R}^2} \textup{d}^2x \, \psi^* \hat{L}_z \psi,
\end{align}
where $g=\tfrac{4\pi \hbar ^2 a_s}{m}$ is known as the short-range interaction parameter, and the angular momentum operator is
\begin{equation}
    \hat{L}_z = -i\hbar \left( x \partial_y - y \partial_x \right).
\end{equation}
There is also a conserved quantity, the number of atoms $N$, which is defined by
\begin{equation}
\label{eq: Number of atoms}
    N = \int_{\mathbb{R}^2} \textup{d}^2 r \, |\psi(\vec{r})|^2.
\end{equation}

The first term in the energy represents the kinetic energy of the condensate, the second is a harmonic trapping potential, and the quartic term encodes short-range interactions between atoms. 
The final term introduces rotation in the $z$-direction, which energetically favors the formation of vortices in the condensate.

The parameter set consists of the particle mass $m$, trapping frequency $\omega$, interaction strength $g$, and rotation frequency $\Omega$, together with $\hbar$. 
In practice, these are often rescaled to a dimensionless form, fixing the normalization $\int |\psi|^2 = N$ and reducing the problem to a small number of effective parameters controlling interaction strength and rotation.

\begin{figure*}[t]
    \centering
    \includegraphics[width=0.7\textwidth]{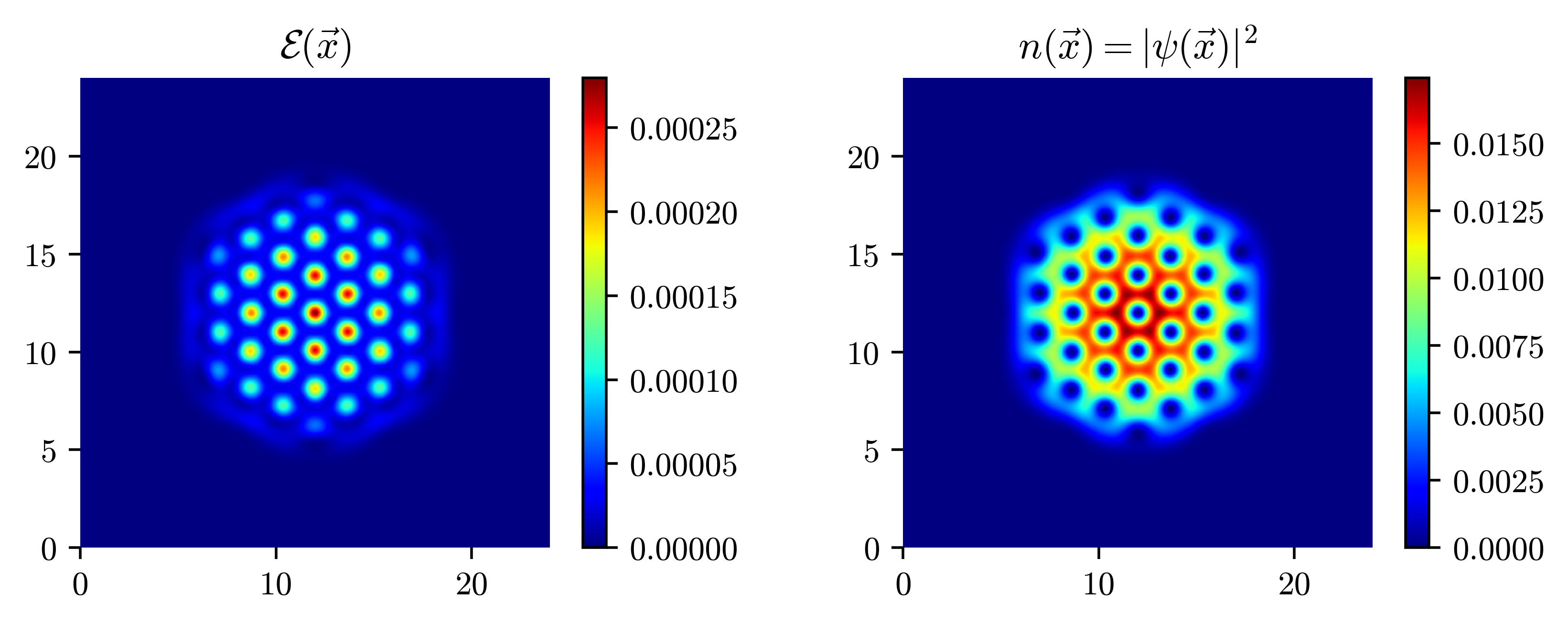}
    \caption{A vortex lattice in a rotating trapped Bose--Einstein condensate. All plots are adimensional. The left panel shows the energy density $\mathcal{E}(\vec{x})$ and the right panel shows the number density $n(\vec{x})$, normalised such that $\int_{\mathbb{R}^2}\textup{d}^2x \, n(\vec{x})=1$. The vortex lattice is obtained in a rotating BEC consisting of $N=1000$ atoms, with atomic mass $m=1.45\times 10^{-25}\,\textup{kg}$. The trap frequency is set to $\omega=200\,\textup{Hz}$ and the superfluid itself is rotating at $\Omega=0.99\omega=198\,\textup{Hz}$. The $s$-wave scattering length is chosen to be $a_s=109a_0$, where $a_0=0.529\times10^{-10}\,\textup{m}$ is the Bohr radius.}
    \label{fig: BEC}
\end{figure*}

In contrast to the other theories considered, the BEC model admits a well-defined analytic ground state in the Thomas--Fermi regime, obtained by neglecting the kinetic term and minimizing the potential energy under a normalization constraint. 
The solver therefore recovers the Thomas--Fermi density profile as the ground state configuration, rather than a non-trivial topological soliton. 
However, when the rotation frequency $\Omega$ is increased, the rotational term drives the nucleation of quantized vortices, and the system transitions from a smooth Thomas--Fermi profile to vortex-carrying states, eventually forming ordered vortex lattices at sufficiently high rotation.
Such a state is plotted in Fig.~\ref{fig: BEC}.

From the implementation perspective, the structure closely parallels the chiral magnet case: the solver minimizes the energy functional subject to the nonlinear dynamics of the field, now without a unit-length constraint but with a conserved particle number. 
The rotational term introduces a non-trivial coupling to angular momentum, leading to vortex nucleation and lattice formation as $\Omega$ is increased.

This example demonstrates the flexibility of the framework in handling complex scalar fields, nonlinear self-interactions, and additional symmetry-breaking terms within the same variational pipeline. 
In particular, it highlights how the same simulation interface used in Fig.~\ref{fig: Usage} extends naturally beyond micromagnetics to superfluid systems, with only the underlying energy functional and parameter set changed.


\section{Impact and reuse potential}
\label{sec:impact}

The primary contribution of \texttt{soliton\_solver} is not a single physical model, but a reusable computational framework for GPU-based simulation of two-dimensional non-linear field theories supporting topological solitons.
Its scientific value lies in the combination of a theory-agnostic CUDA backend, a modular theory registry, model-specific parameter and initialization interfaces, and an interactive GPU-resident visualization pipeline.

This structure supports reuse in several complementary ways.
First, the built-in theory modules can be used directly for studies of vortices, skyrmions, condensate textures, and coupled superconducting--magnetic systems.
Second, new models can be introduced with relatively small additions of code, since the shared solver already provides grid construction, device memory management, finite-difference stencils, Runge--Kutta integration, arrested Newton flow minimization, and reduction operations.
Third, the interactive visualization backend enables exploratory workflows in which initial conditions and metastable configurations can be constructed and refined prior to batch simulation.

The combination of these features makes the package suitable both for production calculations and for rapid prototyping of new models.
In particular, the separation between numerical infrastructure and model specification reduces duplication of effort across related problems and allows researchers to focus on the physics of interest rather than the underlying implementation.


\section{Conclusions}
\label{sec:conclusion}

We have presented \texttt{soliton\_solver}, a GPU-accelerated Python framework for the simulation and real-time visualization of topological solitons in two-dimensional non-linear field theories.
The software combines a reusable finite-difference and minimization backend with a modular theory interface and a CUDA--PyOpenGL rendering pipeline.
This design allows distinct physical models to share a common numerical infrastructure while remaining extensible at the level of theory implementation.

The current package supports multiple built-in theories and is structured so that additional models can be incorporated with minimal modification of the solver core.
This makes it suitable both as a research tool for existing model classes and as a development platform for new GPU-based soliton simulations.

Future work will focus on extending the range of supported theories, improving diagnostics and analysis tools, and further developing the interactive visualization capabilities.
In addition, systematic benchmarking and validation across different model classes will be carried out to assess performance and accuracy in a range of applications.


\section*{Acknowledgments}

The author acknowledges funding from the Olle Engkvists Stiftelse through the grant 226-0103.


\bibliographystyle{elsarticle-num-names} 
\bibliography{main.bib}

\end{document}